\documentclass[reqno]{article}
\newcommand{\ds}{\displaystyle}
\newcommand{\dsf}{\ds\frac}

\newcommand{\beq}{\begin{equation}}
\newcommand{\eeq}{\end{equation}}

\usepackage{ae} 
\usepackage[T1]{fontenc}
\usepackage[ansinew]{inputenc}
\usepackage{amsmath}
\usepackage{graphicx}
\usepackage{color}
\usepackage[colorlinks]{hyperref}
\usepackage{multicol}
\begin{document}
\footnotesize

\begin{center}
\bf Dynamics of the magnetic flux penetration into type II
superconductors
\end{center}

\begin{center}
N. A. Taylanov
\end{center}

\begin{center}
\emph{National University of Uzbekistan}
\end{center}
\begin{center}
{\bf Abstract}
\end{center}
\begin{center}
\mbox{\parbox{13cm}{\footnotesize The magnetic flux penetration
dynamics of type-II superconductors in the flux flow regime is
studied by analytically solving the nonlinear diffusion equation
for the magnetic flux induction, assuming that an applied field
parallel to the surface of the sample and using a power-law
dependence of the differential resistivity on the magnetic field
induction. An exact solution of nonlinear diffusion equation for
the magnetic induction is obtained using a well known self-similar
technique. }}
\end{center}


\begin{multicols}{2}{
\begin{center}
{\bf\S 1. Introduction}
\end{center}

Theoretical investigations of the magnetic flux penetration
dynamics into superconductors in a various regimes with a various
current-voltage characteristics is one of key problems of
electrodynamics of superconductors. Mathematical problem of
theoretical study the dynamics of evolution and penetration of
magnetic flux into the sample in the viscous flux flow regime can
be formulated on the basis of a nonlinear diffusion-like equation
[1-3] for the magnetic field induction in a superconductor [4-14].
The dynamics of space-time evolution of the magnetic flux
penetration into type-II superconductors, where the flux lines are
parallel to the surface of the sample for the viscous flux flow
regime with  a nonlinear relationship between the field and
current density in type II superconductors has been studied by
many authors [4-7]. The magnetic flux penetration problem was
theoretically studied for the particular case, where the flux flow
resistivity independent of the magnetic field by authors [5].
Similar problem has been considered in [6] for the semi-infinite
sample in parallel geometry. The situation, where flux flow
resistivity depends linearly on the magnetic field induction was
considered analytically in [4]. Analogical problem for the creep
regime with a nonlinear relationship between the current and field
has been considered in [8-14]. The magnetic flux penetration into
the superconductor sample, where the flux lines are perpendicular
to the surface of the sample is described by a non-local nonlinear
diffusion equation [7]. This problem has been exactly solved by
Briksin and Dorogovstev [7] for the case thin film geometry in the
flux flow regime of a type-II superconductors.

\vskip 0.5cm
\begin{center}
{\bf \bf\S 2. Objectives}
\end{center}

In the present paper the magnetic flux penetration dynamics of
type-II superconductors in the flux flow regime is studied by
analytically solving the nonlinear diffusion equation for the
magnetic flux induction, assuming that an applied field parallel
to the surface of the sample and using a power-law dependence of
the differential resistivity on the magnetic field induction. An
exact solution of nonlinear diffusion equation for the magnetic
induction $\vec B(r,t)$ is obtained by using a well known
self-similar technique. We study the problem in the framework of a
macroscopic approach, in which all lengths scales are larger than
the flux-line spacing; thus, the superconductor is considered as
an uniform medium.

\vskip 0.5cm
\begin{center}
{\bf\S 3. Formulation of the problem}
\end{center}

Bean [15] has proposed the critical state model which is
successfully used to describe magnetic properties of type II
superconductors. According to this model, the distribution of the
magnetic flux density $\vec B$ and the transport current density
$\vec j$ inside a superconductor is given by a solution of the
equation

\begin{equation}
rot\vec B=\mu_0\vec j.
\end{equation}
When the penetrated magnetic flux changes with time, an electric
field $\vec E(r, t)$ is generated inside the sample according to
Faraday's law

\begin{equation}
rot\vec E=\dsf{d\vec B}{dt}.
\end{equation}
In the flux flow regime the electric field $\vec E(r, t)$ induced
by the moving vortices is related with the local current density
$\vec j(r, t)$ by the nonlinear Ohm's law

\begin{equation}
\vec E=\rho \vec j.
\end{equation}
In combining the relation (3) with Maxwell’s equation (2), we
obtain a nonlinear diffusion equation for the magnetic flux
induction $\vec B(r, t)$ in the following form

\begin{equation}
\dsf{d\vec B}{dt}=\dsf{1}{\mu_0}\nabla\left[\rho(B)\nabla \vec
B\right].
\end{equation}
Formally, this differential equation is simply a nonlinear
diffusion equation with a diffusion coefficient depending on
magnetic induction $B$. The parabolic type diffusion equation (4)
allows to obtain a time and space distribution of the magnetic
induction profile in a superconductor sample. It is evident that
the space-time structure of the solution of the diffusion equation
(4) is determined by the character of dependence of the
differential resistivity coefficient $\rho$ on the magnetic field
induction $B$. Usually, in real experimental situation [16], the
differential resistivity $\rho$ grows with an increase of magnetic
field induction

\begin{equation}
\rho=\dsf{\vec B\phi_0}{\eta c^2}=\rho_n\dsf{\vec B}{H_{c2}},
\end{equation}
where $\rho_n$ is the differential resistivity in the normal
state; $\eta$ is the viscous coefficient, $\phi_0=\pi h c/2e$ is
the magnetic flux quantum, $H_{c2}$ is the upper critical field of
superconductor [16]. In the case, when the differential
resistivity $\rho$ is a linear function of the magnetic field
induction $B$ an exact solution of the diffusion equation (4) can
be easily obtained by using the well-known scaling methods [1, 2].
For the complex dependence of $\rho(B)$ it can be use by empirical
power-law dependence $\rho(B)= B^n$, where n is the positive
constant parameter.

\vskip 0.5cm
\begin{center}
{\bf\S 4. Basic equations}
\end{center}

We formulate the general equation governing the dynamics of the
magnetic field induction in a superconductor sample. We study the
evolution of the magnetic penetration process in a simple geometry
- superconducting semi-infinitive sample $x\geq 0$. We assume that
the external magnetic field induction $B_e$ is parallel to the
z-axis. When the magnetic field with the flux density $B_e(t)$ is
applied in the direction of the z-axis, the transport current
$\vec j(r, t)$ and the electric field $\vec E(r, t)$ are induced
inside the slab along the y-axis. For this geometry, the spatial
and temporal evolution of magnetic field induction $\vec B(r, t)$
is described by the following nonlinear diffusion equation in the
generalized dimensionless form [7]

\begin{equation}
\dsf{db}{dt}=\dsf{d}{dx}\left[b^n\left[\dsf{db}{dx}\right]^q\right],
\end{equation}
where we have introduced the dimensionless parameters $b=B/B_e$,
$j=j/j_c$, $x'=x/x_0$, $t'=t/\tau$  and variables
$x_0=B_e/\mu_0j_{c}$ is the magnetic field penetration depth in a
Bean model; $\tau=\rho_nj_{c}^{2}\mu_0/B_{e}^{2}$ is the
relaxation diffusion time; q is the positive constant parameter.

The diffusion equation (6) can be integrated analytically subject
to appropriate initial and boundary conditions in the center of
the sample and on the sample’s edges. We consider the case, when
the magnetic field applied to sample increases with time according
to a power law with the exponent of $\alpha\geq 0$

\begin{equation}
b(0, t)=b_0t^{\alpha}
\end{equation}
Boundary condition (7) is equivalent to a linear increase in the
magnetic field with time, which corresponds to a real experimental
situation. As can be easily seen that the case $\alpha=0$
describes a constant applied magnetic field at the surface of the
sample, while the case $\alpha=1$ corresponds to linearly
increasing applied field, respectively. The other boundary
condition follows from the continuity of the flux at the free
boundary $x=x_p$

\begin{equation}
b(x_p, t)=0,
\end{equation}
where $x_p$ is the dimensionless position of the front of the
magnetic field. The flux conservation condition for the magnetic
field induction can be formulated in the following integral form

\begin{equation}
\int b(x, 0)dx=1.
\end{equation}
It should be noted that the nonlinear diffusion equation (6),
completed by the boundary conditions for magnetic induction,
totally determines the problem of the space-time distribution of
the magnetic flux penetration into superconductor sample in the
flux flow regime with a power-law dependence of differential
resistivity on the magnetic field induction. Solution of this
equation gives a complete description of the time and space
evolution of the magnetic flux in a sample.

\vskip 0.5cm
\begin{center}
{\bf\S 5. Scaling solution}
\end{center}

In the following analysis we derive an evolution equation for the
magnetic induction profile and formulate a similarity solution for
the b(x, t). As can be shown that the nonlinear diffusion equation
(6) can be solved exactly using well known scaling methods [1, 2].
At long times we present a solution of the nonlinear diffusion
equation for the magnetic induction (6) in the following scaling
form

\begin{equation}
b(x, t)=t^{\alpha}f\left(\dsf{x}{t^{\beta}}\right).
\end{equation}
The similarity exponents $\alpha$ and $\beta$ are of primary
physical importance since the parameter $\alpha$ represents the
rate of decay of the magnetic induction b(x, t), while the
parameter $\beta$ is the rate of spread of the space distribution
as time goes on. Inserting this scaling form into differential
equation (6) and comparing powers of t in all terms, we get the
following relationship for the exponents $\alpha$ and $\beta$
$$
\alpha+1=\alpha(n+q)+\beta(1+q).
$$
Using the condition of the flux conservation (9) we obtain
$$
\alpha=\beta=-\dsf{1}{n+2q}
$$
which suggests the existence of self-similar solutions in the form

\begin{equation}
b(z)=t^{-1/(n+2q)}f(z),\quad z=x/t^{1/(n+2q)}.
\end{equation}
Substituting this scaling solution (11) into the governing
equation (6) yields an ordinary differential equation for the
scaling function f(z) in the form
\begin{equation}
(n+2q)\dsf{d}{dz}\left[f^n\left(\dsf{df}{dz}\right)^q\right]+z\dsf{df}{dz}+f=0.
\end{equation}
The boundary conditions for the function $f(z)$ now become

\begin{equation}
f(0)=1, \quad f(z_0)=0.
\end{equation}
The above equation (12), depending on the initial and the boundary
conditions describes a scaling—like behavior magnetic flux front
with a time—dependent velocity in the sample. After a further
integration and applying the boundary conditions (13) we get the
following solution of the problem

\begin{equation}
f(z)=f(z_0)\left[1-\left(\dsf{z}{z_0}\right)^{1+q}\right]^{1/(n+q-1)}.
\end{equation}
$$
f(z_0)=\left[\dsf{n+q-1}{1+q}\left(\dsf{1}{n+2q}\right)^{1/q}\right]^{q/(n+q-1)}
z_{0}^{(q+1)/(n+q-1)}.
$$
The position of the front $z_0$ can now be found by substituting
the solution (14) into the integral condition (9) and it is given
by

\begin{equation}
z_{0}^{(n+2q)/(n+q-1)}=\dsf{\left[\dsf{n+q-1}{1+q}\left(\dsf{1}{n+2q}\right)^{1/q}\right]^{q/(n+q-1)}}{
\dsf{1}{q+1}\dsf{\Gamma\left(\dsf{n+q}{n+q-1}+\dsf{1}{2}\right)}
{\Gamma\left(\dsf{n+q}{n+q-1}\right)\Gamma\left(\dsf{1}{q+1}\right)}}.
\end{equation}
It is convenient to write the self-similar solution (14) in terms
of a primitive variables, as

\begin{equation}
b(x,
t)=b_0(t)\left[1-\left(\dsf{x}{x_p}\right)^{1+q}\right]^{1/(n+q-1)}.
\end{equation}
$$
b_0(t)=t^{-1/(n+2q)}\left[\dsf{n+q-1}{1+q}\left(\dsf{z_{0}^{(q+1)/q}}{n+2q}\right)^{1/q}\right]^{q/(n+q-1)}.
$$
This solution describes the propagation of the magnetic field into
the sample, the magnetic induction being localized in the domain
between the surface x=0 and the flux front $x_p$. This solution is
positive in the plane $x_{p}^{2}>x^2$ and is zero outside of it.
Note, that only the $x>0$ and $t>0$ quarter of the plane is
presented, because of it has physical relevance. The penetrating
flux front position  $x=x_p(t)$ as a function of time can be
described by the relation

$$
v\sim\dsf{dx_p}{dt}\sim t^{-(2q+n-1)/(n+2q)}.
$$

\begin{center}
\includegraphics[width=2.5583in]{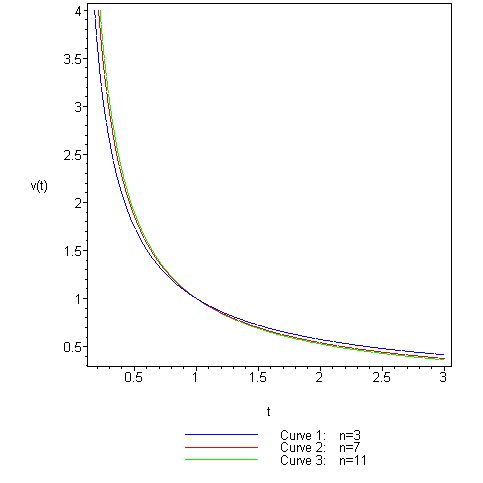}
\end{center}
\begin{center}
Fig.1. The profile of the magnetic flux front velocity at different values of n=3, 7, 11.\\
\end{center}

The velocity of the magnetic flux front decreases rapidly as the
magnetic flux propagates (Fig1).

\vskip 3cm
\begin{center}
{\bf\S 6. Particular case}
\end{center}

The spatial and temporal profiles of magnetic flux penetration in
the sample depends on the set of three independent parameters, n,
q and $\alpha$. It is of interest to consider the nonlinear
diffusion equation for the magnetic induction for different values
of the exponents n, q and $\alpha$. For a given parameter set n, q
and $\alpha$ the form of the scaling function f(z) can be obtained
by solving the nonlinear diffusion equation (6) analytically by a
self-similar technique. We solve the nonlinear diffusion equation
analytically to provide expressions for the time-space evolution
of the magnetic induction for different values of exponents n, q
and $\alpha$. Next, we systematically analyze the effect of
different values of exponents on the shape of the magnetic flux
front in the sample. Varying the parameters of the equation, we
may observe a various shapes of the magnetic flux front in the
sample. A similar approach has been presented in Ref. [7] within
the framework of non-linear flux diffusion in transverse geometry.
As can be shown below that different values exponents n and q
generate different space–time magnetic flux fronts. Below we
consider a few more practically relevant examples for which the
magnetic flux front has a different shape depending on the
different values of exponents n and q.

\vskip 0.5cm
\begin{center}
{\bf\S 6. 1.  Case $q=1$ }
\end{center}

Let us first consider the most interesting case q=1. In this
particular case the spatial and temporal evolution of the magnetic
flux induction is totally determined by the parameters n and
$\alpha$. In the following analysis we derive an evolution
equation for the magnetic induction profile and apply the scalings
of the previous section to formulate a similarity solution for the
b(x, t). For this particular case nonlinear diffusion equation (6)
can be solved exactly using the scaling method. Thus, based on the
scalings described in the previous section, we get the following
relation for the exponents

$$
\alpha=\beta=-\dsf{1}{n+2}.
$$
The last relation suggests the existence of solution to equation
(6) of the form
\begin{equation}
b(x, t)=t^{-1/(n+2)}f(z),\quad z=x/t^{1/(n+2)}.
\end{equation}
Substituting the similarity solution (17) into the governing
equation (6) yields an ordinary differential equation for the
scaling function f(z)

\begin{equation}
(2+n)\dsf{d}{dz}\left(f^n\dsf{df}{dz}\right)+z\dsf{df}{dz}+f=0.
\end{equation}
Integrating the equation (18) by parts and applying the boundary
conditions (13) give

\begin{equation}
f(z)=\left[\dsf{n}{2(n+2)}z_{0}^{2}\right]^{1/n}
\left[1-\dsf{z^2}{z_{0}^{2}}\right]^{1/n},
\end{equation}
which is the explicit form of the similarity solution, which we
have been seeking. The position of the front $z_0$ can now be
found by substituting the last solution into the integral
condition (9), so we have

$$
\left[\dsf{n}{2(n+2)}z_{0}^{2}\right]^{1/n}
\int_{0}^{z_0}\left[1-\dsf{z^2}{z_{0}^{2}}\right]^{1/n}dz=1,
$$
By using the following transformation
$$
z=z_0\sin \omega,
$$
and after integrating we obtain
$$
z_{0}^{(n+2)/n}\left[\dsf{n}{2(n+2)}\right]^{1/n}=\dsf{2}{\sqrt{\pi}}\dsf{\Gamma\left(\dsf{3}{2}+\dsf{1}{n}\right)}
{\Gamma\left(1+\dsf{1}{n}\right)}.
$$
It is convenient to write the self-similar solution (19) in terms
of a primitive variables, as
\begin{equation}
b(x, t)=b_0\left[1-\dsf{x^2}{x_{p}^{2}}\right]^{1/n},
\end{equation}
where
$$
b_0=\left[\dsf{n}{2(n+2)}z_{0}^{2}\right]^{1/n}t^{-1/(n+2)}.
$$
Equation (20) constitutes an exact solution of the nonlinear
flux-diffusion equation for the situation, when q=1. As can be
seen the solution (20) describes the propagation of the flux
profile inside the sample. The profile of the the normalized flux
density $b(x, t)$ for this case is shown schematically in figure
2.

\begin{center}
\includegraphics[width=2.5583in]{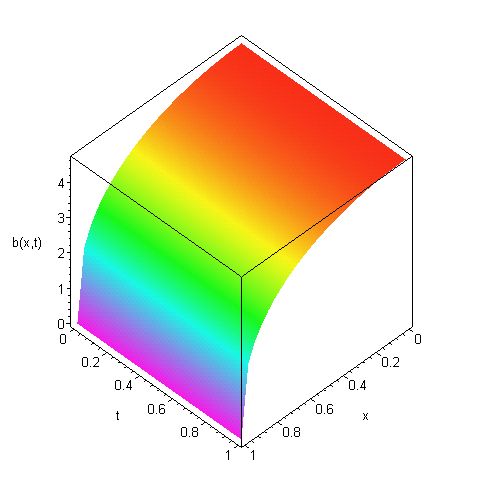}

Fig.2. The distribution of the normalized flux density $b(x, t)$
at different times t=0.1, 0.2, 0.3 for n=1, q=1.
\end{center}

The penetrating flux front position $x=x_p(t)$ as a function of
time can be described by the relation
$$
x_p=z_0t^{1/(n+2)}.
$$
The velocity of penetration of a magnetic flux into a
superconductor can be naturally determined from the last relation
$$
v\sim t^{-{(n+1)}/(n+2)}.
$$
Interestingly, that the normalized current density $j(x,t)$ in the
region, $0<x<x_p$ can be presented using the equations
$$
j(x, t)=-\dsf{1}{c}\dsf{db}{dx},
$$
After a simple analytical calculation, we can easily obtain the
space and time profiles of the normalized current density $j(x,t)$
 in the following form
$$
j(x,t)=\dsf{2b_0}{cnx_{p}^{2}}\left[1-\left(\dsf{x}{x_p}\right)^2\right]^{1/n-1}x,
$$

 \vskip 0.5cm
\begin{center}
{\bf\S 6. 2.  Case $n=0$ }
\end{center}

Let us consider the case $n=0$. For this particular case the
long-time asymptotic behavior of the magnetic induction has the
scaling form

\begin{equation}
b(x, t)=f(z),\quad z=xt^{-1/(q+1)}.
\end{equation}

Substituting the scaling solution into the governing equation (6)
yields an ordinary differential equation for for the distribution
f(z) in the form
$$
(q+1)\dsf{d}{dz}\left[\dsf{df}{dz}\right]^q+z\dsf{df}{dz}=0.
$$
Integrating the last equation and applying the boundary conditions
(13) we get the following solution of the problem

\begin{equation}
j(z)=-\dsf{df}{dz}=\left[\dsf{q-1}{2q(q+1)}(z_{0}^{2}-z^2)\right]^{1/(q-1)},
\end{equation}
where the position of the front $z_0$ can be found by substituting
the solution (22) into the integral condition (9), and it is given
by the following expression

$$
z_0=\left[2q\dsf{q+1}{q-1}\left(\dsf{\sqrt{\pi}}{2}\dsf{\Gamma\left(\dsf{q}{q-1}
\right)}{\Gamma\left(\dsf{3}{2}+\dsf{1}{q-1}\right)}\right)^{1-q}\right]^{1/(q+1)}.
$$
The scaling solution (22) can be written in terms of the primitive
variables as
$$
j(x,
t)=\dsf{1}{t^{1/(q+1)}}\left[\dsf{q-1}{2q(q+1)}\right]^{1/(q-1)}z_{0}^{2/(q-1)}
\left[1-\dsf{x^2}{x_{p}^{2}}\right]^{1/(q-1)}.
$$
The the flux front can be approximately given as
$x_p=z_0t^{1/(q+1)}$. The velocity of penetration of a current
density into a superconductor is determined from the relation

$$
v=\dsf{dx_p}{dt}\sim t^{-q/(q+1)}.
$$
We note that, an analogous problem has also been studied in [10,
11] in connection with the magnetic relaxation of a
superconducting slab in the flux creep regime in the framework of
an approximate power-law dependence of the electric field E on the
current density j. The authors [11] showed that in the case
logarithmic barriers the relaxation process causes the system
self-organize into critical state. Supposing that the homogeneous
magnetic induction $B_0$ is induced by a constant magnetic field,
they found the expression for the magnetization moment in the
limit of $n\gg 1$. It has been shown by Koziol [12] that a pinning
potential depending logarithmicaly on current density leads to a
similar nonlinear diffusion equation for the spatiotemporal
evolution of the flux density with a a power-law current-voltage
characteristic. An approximate and exact solutions of the
diffusion problem have been derived assuming that an external
magnetic field directed to parallel of the surface of a sample.
The scaling relation between the characteristic relaxation time,
magnetic field and a sample size has been found. Similar problem
has been considered by Gilchrist [8, 9]. The flux creep problem
has been solved by Wang et.all., [13] and for an exponential model
by authors [14], numerically.

\vskip 0.5cm
\begin{center}
{\bf\S 6. 3.  Case $n=1$ }
\end{center}

Let us now consider the case n=1. In this particular case the
spatial and temporal evolution of the magnetic flux induction is
determined by the parameters q and $\alpha$. In the following
analysis we derive an evolution equation for the magnetic
induction profile for the case n=1 and apply the scalings of the
previous section to formulate a similarity solution for the b(x,
t). For this particular case nonlinear diffusion equation (6) can
be solved exactly using the scaling method. Thus, based on the
scalings described in the previous section, we get the following
relation for the exponents

$$
\alpha=\beta=-\dsf{1}{2q+1}.
$$
The last relation suggests the existence of solutions of the form
\begin{equation}
b(x, t)=t^{-1/(2q+1)}f(z),\quad z=x/t^{1/(2q+1)}.
\end{equation}

Substituting the similarity solution (23) into the governing
equation (6) yields an ordinary differential equation for the
function f(z)

\begin{equation}
(2q+1)\dsf{d}{dz}\left(f\left(\dsf{df}{dz}\right)^q\right)+z\dsf{df}{dz}+f=0.
\end{equation}
Integrating the equation (24) by parts and applying the boundary
conditions give

\begin{equation}
f(z)=\left(\left(\dsf{q}{q+1}\right)^q\dsf{z_{0}^{q+1}}{2q+1}\right)^{1/q}
\left[1-\left(\dsf{z}{z_0}\right)^{(q+1)/q} \right],
\end{equation}
which is the explicit form of the similarity solution we have been
seeking. The position of the front can now be found by
substituting the last solution into the integral condition (9),
and we have

$$
z_0=[q^{1/(q+1)}(2q+1)])^{(q+1)/(2q+1)}.
$$
It is convenient to write the self-similar solution (25) in terms
of a primitive variables, as
\begin{equation}
b(x, t)=b_0\left[1-\dsf{x^2}{x_{p}^{2}}\right]^{1/q},
\end{equation}
where
$$
b(x,
t)=t^{-1/(2q+1)}\left(\left(\dsf{q}{q+1}\right)^q\dsf{z_{0}^{q+1}}{2q+1}\right)^{1/q}.
$$
Equation (26) describes an exact solution of the nonlinear
flux-diffusion equation for the situation, when n=1. As can be
seen the solution (26) describes the propagation of the flux
profile into the sample. The the flux front can be approximately
given as $x_p=t^{1/(2q+1)}$. The velocity of penetration of a
magnetic flux induction front into a superconductor is determined
from relation
$$
v=\dsf{dx_p}{dt}\sim t^{-2q/(2q+1)}.
$$
The velocity of the magnetic flux front decreases rapidly as the
magnetic flux propagates.

\vskip 0.5cm
\begin{center}
{\bf\S 6. 4.  Case $n=1, \quad q=1$}
\end{center}

Let us now consider the case, where $n=1, \quad q=1$. For this
particular case the nonlinear diffusion equation for the
distribution magnetic flux admits an exact self-similar solution
with a sharp front moving with a constant velocity. The scaling
analysis show that the diffusion equation can be solved for the
following values of parameters 1). $\alpha$=0, $\beta$=1/2; \quad
2). $\alpha$=1/3, $\beta$=2/3; \quad 3). $\alpha$=1, $\beta$=1;
\quad 4). $\alpha$=1/3, $\beta$=1/3. Let us consider a solution of
the nonlinear diffusion problem for the exponents $\alpha$=1/3,
$\beta$=1/3. In this case the nonlinear diffusion equation (6)
admits an exact solution [4] with the similarity variable

\begin{equation}
b(x, t)=t^{-1/3}f(xt^{-1/3}).
\end{equation}
By substituting the solution (27) into diffusion equation (6) we
obtain an ordinary differential equation for the scaling function
f(z) in the form

\begin{equation}
3\dsf{d}{dz}\left(f\dsf{df}{dz}\right)+z\dsf{df}{dz}+f=0.
\end{equation}
Integrating the equation (28) and using the boundary conditions
for the magnetic induction $f(z)$ we get a simple equation

\begin{equation}
3\dsf{df}{dz}+z=0.
\end{equation}
Separating variables and integrating equation (29) gives an
implicit solution in the form

\begin{equation}
f=\dsf{1}{6}(z_{0}^{2}-z^2),
\end{equation}
where the integration constant is determined by integral relation
(9) and has the form

$$
z_0=\left(\dsf{9}{2}\right)^{1/3}.
$$
Thus, we have solution for the magnetic induction in a primitive
variables

\begin{equation}
b(x,
t)=\dsf{1}{6t^{1/3}}\left[\left(\dsf{9}{2}\right)^{2/3}-\dsf{x^2}{t^{2/3}}\right].
\end{equation}
The expression (31) describes an exact solution of the nonlinear
flux-diffusion equation for the situation, when $n=1, \quad q=1$.
As can be seen the solution (31) describes the propagation of the
flux profile into the sample. The profiles of the the normalized
flux density $b(x, t)$ and electric field density $e(x, t)$ for
this case is shown schematically in the figures 3a-3b.

\begin{center}
\includegraphics[width=2.5583in]{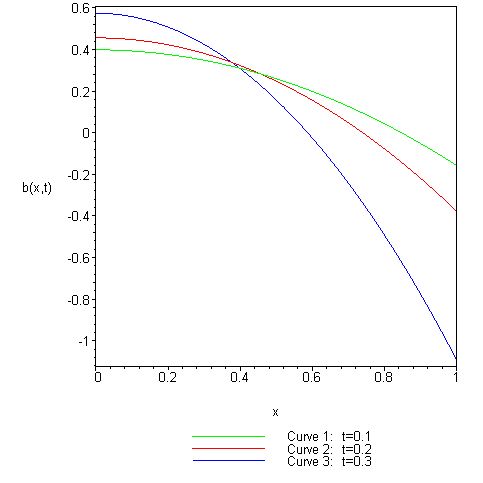}
\includegraphics[width=2.5583in]{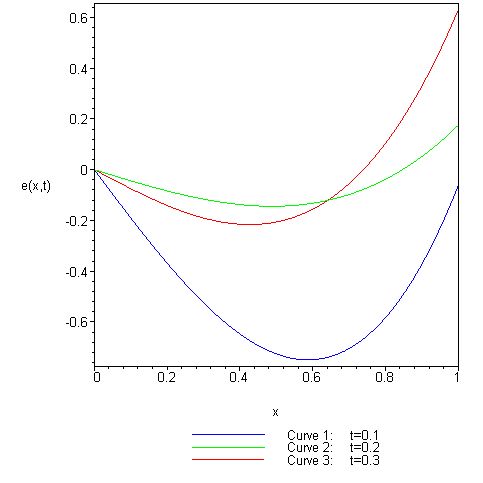}\\
Figs.3a and 3b. The distributions of the normalized flux density
$b(x, t)$ and electric field density $e(x, t)$ at different times
t=0.1, 0.2, 0.3 for   n=1, q=1.
\end{center}

It is evident from the solution (31) that the shock wave like
magnetic flux propagates with wave front

\begin{equation}
x_p=z_0t^{1/3}.
\end{equation}

It can be shown that the sharp flux front moves with the time
dependent velocity

\begin{equation}
v=\dsf{dx_p}{dt}\sim z_0t^{-2/3}.
\end{equation}
The case $\alpha$=1/3, $\beta$=1/3 implies that the flux front
asymptotically expands according to power-law $x_p\propto t^{1/3}$
and magnetic flux penetration profile decreases as $b\propto
t^{-1/3}$. Similar problem for the case penetration of magnetic
field into superconductors has been studied by Bass [4]. An
analogous theoretical result in the viscous flow mode for vortices
was obtained in [5] using the model of the critical state for
oxide high-$T_c$ superconductors.

Finally, in the case $\alpha$=0, $\beta$=1/2 the flux front
position has the form of $x_p\propto t^{1/2}$ and an approximate
analytical solution for the case can be easily found solving the
diffusion equation with the help of similarity variable $b(x,
t)=f(z)$, where $z=f(x/t^{1/2})$. In the vicinity of the front
($x\propto x_p$) the magnetic flux profile is given  by the
following simple relation $b\propto\sqrt{2(x-x_p)}$. In this case,
which corresponds to a constant external magnetic field the flux
front position $x_p$ grows with time as $x_p\propto t^{1/2}$ for
long times.

\vskip 0.5cm
\begin{center}
{\bf\S 6. 5.  Case $n=2, \quad q=1$ }
\end{center}

The scaling analysis show that the diffusion equation can be
solved for the following values of parameters 1)$\alpha$=1/4,
$\beta$=1/4; \quad 2) $\alpha$=1/4, $\beta$=1/6. The solution to
diffusion equation for this case can be presented in the following
form

\begin{equation}
b(x, t)=t^{-1/4}f(x/t^{1/4}).
\end{equation}
Substituting the solution (34) into the partial differential
equation (6) and integrating with the boundary conditions we can
see that magnetic flux profile has an exact solution in the form

$$
f=\dsf{z_0}{2}\left[1-\dsf{z^2}{z_{0}^{2}}\right]^{1/2},
$$
where the integration constant has the form
$$
z_0=\dsf{2}{\sqrt{\pi}}.
$$
The solution for the magnetic induction in a primitive variables
has the form

\begin{equation}
b(x,t)=\dsf{1}{\sqrt{\pi
}t^{1/4}}\left[1-\dsf{\pi}{4}\dsf{x^2}{t^{1/2}}\right]^{1/2}.
\end{equation}
The expression (35) describes an exact solution of the nonlinear
flux-diffusion equation for the situation, when $n=2, \quad q=1$.
As can be seen the solution (35) describes the propagation of the
flux profile into the sample. The evolution of the self-simulating
process of magnetic field penetration into a superconductor is
shown schematically in the figures 4a-4b.

\begin{center}
\includegraphics[width=2.5583in]{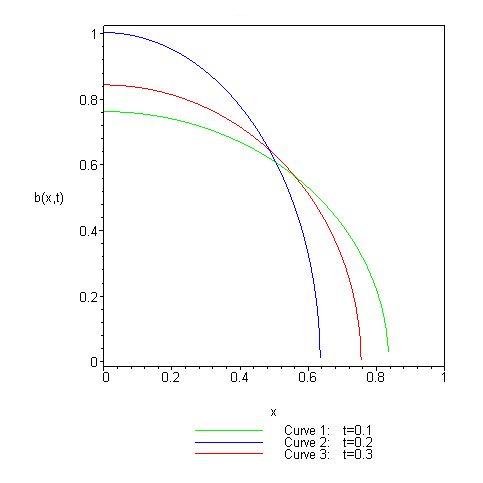}
\includegraphics[width=2.5583in]{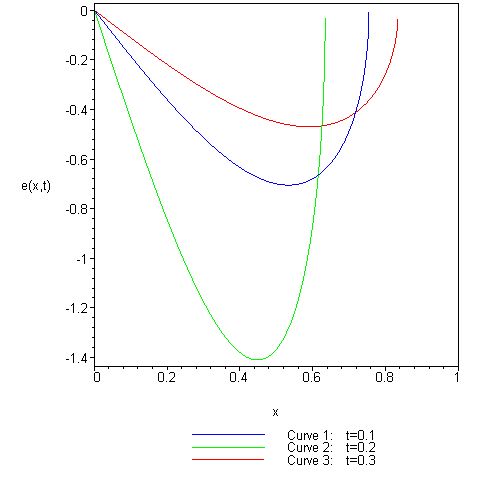}\\
Figs.4a and 4b. The distributions of the normalized flux density
$b(x, t)$ and electric field density $e(x, t)$ at different times
t=0.1, 0.2, 0.3 for $n=2, \quad q=1$.
\end{center}

In this case the front coordinate is $x_p=t^{1/4}$. A main result
is that the magnetic flux will move as a shock front with velocity
$v=t^{3/4}$. Similar problem for the case penetration of magnetic
field into superconductors has been studied in [5].

\vskip 0.5cm
\begin{center}
{\bf\S 6. 6. Case $n=0, \quad q=1$ }
\end{center}

In this case the diffusion equation can be solved in a different
values of $\alpha$ and $\beta$, in particular 1). $\alpha$=0,
$\beta$=1/2; 2). $\alpha$=-1/2, $\beta$=-1/2; 3). $\alpha$=1,
$\beta$=1. Let us consider the case $\alpha$=0, $\beta$=-1/2. In
this case the differential equation is linear and admits an exact
analytical solution. At long times we present a solution of the
nonlinear diffusion equation for the magnetic induction (6) in the
following scaling form

\begin{equation}
b(x, t)=t^{\alpha}f\left[\dsf{x}{t^{\beta}}\right].
\end{equation}
Inserting the scaling form (36) into differential equation (6) we
have
\begin{equation}
t^{\alpha+2\beta}\dsf{d^2f}{dz^2}=t^{\alpha-1}\left(\alpha f+\beta
z\dsf{df}{dz}\right).
\end{equation}
Comparing powers of t in all terms in (37), we get the following
relationship for the exponents $\alpha$ and $\beta$

$$
\alpha=\beta=-\dsf{1}{2},
$$
Then the problem admits a similarity solution of the form

\begin{equation}
b=t^{-1/2}f(xt^{-1/2}).
\end{equation}
Substituting the expression (38) into the nonlinear diffusion
equation (6) gives an ordinary differential equation for the
function similarity function f(z), namely

\begin{equation}
2\dsf{d^2f}{dz^2}+z\dsf{df}{dz}+f=0.
\end{equation}
Integrating  (39) twice and applying the boundary conditions we
have a self-similar solution in the form

\begin{equation}
b(x, t)=\dsf{1}{\sqrt{4\pi
t}}\exp\left[-\dsf{x^2}{x_{p}^{2}}\right].
\end{equation}
where
$$
x_p=\sqrt{4t}.
$$
The last formulae describes an exact solution of the nonlinear
flux-diffusion equation for the situation, when $n=2, \quad q=1$.
As can be seen this solution describes the propagation of the flux
profile into the sample. The profile of the self-simulating
process of magnetic field penetration into a superconductor for
this case is shown schematically in the figures 5a-5b.

\begin{center}
\includegraphics[width=2.5583in]{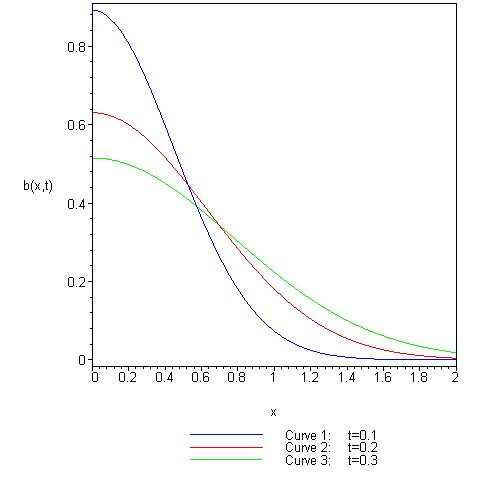}
\includegraphics[width=2.5583in]{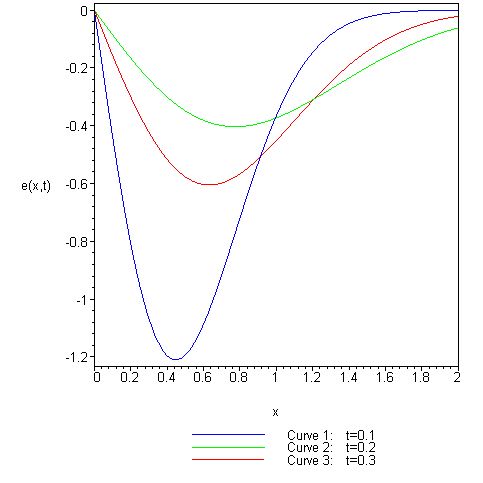}\\
Figs.5a and 5b. The distributions of the normalized flux density
$b(x, t)$ and electric field density $e(x, t)$ at different times
t=0.1, 0.2, 0.3 for $n=0, \quad q=1$ .
\end{center}

Similar problem has been solved in [6] using an extended critical
state model for type-II superconductors. Assuming that the
critical current density and resistivity are constants a
one-dimensional diffusion equation for the magnetic field were
solved analytically for the case when an external magnetic field
is increased with time according to a power law with the exponent
of one- half.

\vskip 0.5cm
\begin{center}
{\bf\S 7. Flux creep}
\end{center}
Let us consider the magnetic flux penetration process for the flux
creep regime. According to Kim-Anderson theory [17, 18] the
thermally activated flux motion in superconductor sample is
described by an Arrhenius-type expression

\begin{equation}
v=v_0 \exp[-U/kT],
\end{equation}
where $v_0$ is the resistivity at T=0, $U$ is an activation energy
for thermally activated flux jumps, and determines the vortex
pinning; $T$ is the temperature and $k$ is the Boltzmann constant.
The activation energy $U=U(\vec j, \vec B, T)$ depends on
temperature $T$, magnetic field induction $\vec B$ and current
density $\vec j$. For the simple case it can be presented by
Kim-Anderson [17] formulae

\begin{equation}
U(j)=U_0\left(1-\dsf{j}{j_c}\right),
\end{equation}
here $U_0$ is the characteristic scale of the activation energy
and $j_c=j_c(B)$ is the critical current density. In the flux
creep state the effective activation energy $U$ grows
logarithmically [19] with decreasing current density as

\begin{equation}
U(j)=U_0 \ln\left(\dsf{j_c}{j}\right)^n,
\end{equation}
where the exponent $n$ depends upon the flux creep regime. The
expression (43) gives a quite realistic description for activation
barriers in a wide range of temperatures and magnetic fields. The
power law characteristic dependence for $U(\vec j)$ has been
observed in numerous experiments, and it has been used extensively
in recent theoretical studies for the field and current
distributions in superconductors [19, 20]. Equation (43) is
equivalent to a power law for the flux-flow resistivity

\begin{equation}
v=v_0\left|\dsf{j}{j_c}\right|^n.
\end{equation}
In this case the phenomenological relation $\vec E(\vec j)$ may be
chosen in the power-law form

\begin{equation}
\vec E=v_0|B|\left|\dsf{\vec j}{j_c}\right|^n\cdot \vec j.
\end{equation}
If $n=1$ the last equation reduces to Ohm's law, describing the
normal or flux-flow state. For infinitely large $n$, the equation
describes the Bean critical state model $j=j_c$ [15]. When
$1<n<\infty$, the last equation describes nonlinear flux creep.

In such case the analytic solution of the nonlinear creep equation
can be constructed by choosing the critical current dependence on
the magnetic field. Many models have been for the functional form
of $j_c(B)$.  For the critical current we adopt the power-law
model [21], which can be applied over a relatively wide magnetic
field range except in the high field region near the upper
critical field

\begin{equation}
j_c(B)=j_0\left(\dsf{B_0}{B}\right)^{\gamma}.
\end{equation}
where $j_0$ and $B_0$ are the characteristic values of the current
density and magnetic field induction; $\gamma$ is the
dimensionless pinning parameter, usually $0<\gamma<1$. If we
assume $\gamma$=0, the above model reduces to the Bean-London
model [15]. This model is applicable to the case where $j_c$ can
be regarded approximately field independent. This power-law model
for the critical current has also been used by other groups [14].
Another possible decay law would be exponential, which has often
been used to take into account its decrease with the magnetic
field [22]

$$
j=j_0\exp{\left(-\dsf{B}{B_0}\right)},
$$
where $B_0$ is a phenomenological parameter related to the pinning
ability: the smaller it is, the more drastic is the decrease of
the critical current with field. The numerical methods have been
applied to resolve the flux diffusion equation, employing the
exponential critical state model [14].

Next, based on the power law and exponential models, we shall
study the distribution of the magnetic induction, current density
and magnetization of superconductors.

\vskip 0.5cm
\begin{center}
{\bf\S 7.1. Power law model}
\end{center}
For the one-dimensional geometry the spatial and temporal
evolution of magnetic field induction $\vec B(r, t)$ is described
by the following nonlinear diffusion equation [10] in the
dimensionless form

\begin{equation}
\dsf{db}{dt}=\dsf{d}{dx}\left[b^{\gamma
n+1}\left|\dsf{db}{dx}\right|^{n-1} \dsf{db}{dx}\right],
\end{equation}
where we have introduced the dimensionless variables and
parameters

$$
b=\dsf{B}{B_0},\quad x_p=\dsf{\mu_0j_{0}}{B_0}x,\quad
t=\dsf{t}{\tau},\quad j=\dsf{j}{j_{0}},
$$
$$
b=\epsilon=\dsf{E}{v_0B_0}, \quad B_0=\mu_0j_{0}v_0\tau.
$$
The solution of above parabolic type diffusion equation allows to
obtain the time and space distribution of the magnetic induction
profile in the considered sample. Inserting the scaling form (10)
into differential equation (47) and comparing powers of t in all
terms, we get the following relationship for the exponents
$\alpha$ and $\beta$

\begin{equation}
\alpha+1=\beta+\alpha(\gamma n+1)+\alpha n+\beta n.
\end{equation}
Using the condition of the flux conservation we obtain

$$
\alpha=\beta=\dsf{1}{(2n+n\gamma+1)}
$$
which suggests the existence of self-similar solutions in the form

\begin{equation}
b(x, t)=t^{-1/(2n+n\gamma+1)}f(z),\quad z=x/t^{1/(2n+n\gamma+1)}.
\end{equation}
Substituting this scaling solution (49) into the governing
equation (47) yields an ordinary differential equation for the
scaling function f(z) in the form

\begin{equation}
\dsf{d}{dz}\left[f^{\gamma
n+1}\left|\dsf{df}{dz}\right|^n\right]+\dsf{1}{(2n+n\gamma+1)}\dsf{d}{dz}\left(z\dsf{df}{dz}\right)=0.
\end{equation}
The solution of the equation (50) must satisfy the following
boundary conditions for the scaling function f(z)

\begin{equation}
f(0)=1, \quad f(z_0)=0.
\end{equation}
After further integrating of equation (50) and applying the
boundary conditions we get the following solution of the problem

\begin{equation}
f(z)=f(z_0)\left[
1-\left(\dsf{z}{z_0}\right)^{(n+1)/n}\right]^{1/(\gamma+1)}
\end{equation}
$$
f(z_0)=\left[n\dsf{(\gamma+1)}{(n+1)}\left(\dsf{z_{0}^{(n+1)}}{2n+n\gamma+1}\right)^{1/n}\right]^{1/(\gamma+1)}.
$$
The position of the front $z_0$ can now be found by substituting
the solution (52) into the integral condition (9) and it is given
by

$$
z_{0}^{(2n+n\gamma+1)/(\gamma+1)}=\dsf{\dsf{n}{n+1}
\dsf{F\left(\dsf{\gamma+2}{\gamma+1}+\dsf{1}{2}\right)}{\Gamma\left(\dsf{\gamma+2}
{\gamma+1}\right)\Gamma\left(\dsf{n}{n+1}\right)}}{\left[n\dsf{(\gamma+1)}{(n+1)}
\left(\dsf{1}{2n+n\gamma+1}\right)^{1/n}\right]^{1/(\gamma+1)}}.
$$
The last solution can be presented as

\begin{equation}
b(x, t)=b_0\left[
1-\left(\dsf{x}{x_p}\right)^{(n+1)/n}\right]^{1/(\gamma+1)}
\end{equation}
where
$$
b_0(0,
t)=t^{-1/(2n+n\gamma+1)}\left[n\dsf{(\gamma+1)}{(n+1)}\left(\dsf{z_{0}^{(n+1)}}{2n+n\gamma+1}\right)^{1/n}
\right]^{1/(\gamma+1)}.
$$
This solution describes the propagation of the magnetic field into
the sample, the magnetic induction being localized in the domain
between the surface x=0 and the flux front $x_p$. The flux front
can be approximately given as $x_p=s_0t^{1/(2n+n\gamma+1)}$. The
velocity of penetration of the magnetic flux induction front into
the superconductor sample is determined from the relation

\begin{equation}
v\sim t^{-n(2+\gamma)/(2n+n\gamma+1)}.
\end{equation}
The velocity of the magnetic flux front decreases rapidly as the
magnetic flux propagates.

Let us consider the most interesting case n=1. In this particular
case the spatial and temporal evolution of the magnetic flux
induction is totally determined by the parameters $\gamma$,
$\alpha$ and $\beta$. In the following analysis we derive an
evolution equation for the magnetic induction profile for the case
n=1 and apply the scalings of the previous section to formulate a
similarity solution for the b(x, t). Based on the scalings
described in the previous section, we get the following relation
for the exponents

\begin{equation}
\alpha=\beta=\dsf{1}{\gamma+3}.
\end{equation}
The last relation suggests the existence of solutions of the form

\begin{equation}
b(x, t)=t^{-1/(\gamma+3)}f(z),\quad z=x/t^{1/(\gamma+3)}.
\end{equation}
Substituting the similarity function (56) into the governing
equation (47) yields the following solution of the problem

\begin{equation}
f(z)=\left[\dsf{z_{0}^{2}}{2}\dsf{(\gamma+1)}{(\gamma+3)}\right]^{1/(\gamma+1)}
\left[1-\left(\dsf{z}{z_0}\right)^2\right]^{1/(\gamma+1)},
\end{equation}
which is the explicit form of the similarity solution we have been
seeking. The position of the front has the form

\begin{equation}
z_{0}^{(\gamma+3)/(\gamma+1)}=\left[\dsf{2}{1}\dsf{(\gamma+3)}{(\gamma+1)}\right]^{1/(\gamma+1)}
\dsf{2}{\sqrt{\pi}}\dsf{G\left(\dsf{1}{1+\gamma}+\dsf{3}{2}\right)}{G\left(\dsf{1}{1+\gamma}+1\right)}.
\end{equation}
or
\begin{equation}
b(x, t)=b_0\left[1-\dsf{x^2}{x_{p}^{2}}\right]^{1/(\gamma+1)},
\end{equation}
where
$$
b_0=\left[\dsf{z_{0}^{2}}{2}\dsf{(\gamma+1)}{(\gamma+3)}\right]^{1/(\gamma+1)}t^{-1/(\gamma+3)}.
$$
Equation (59) constitutes an exact solution of the nonlinear
flux-diffusion equation for the situation, when n=1. The evolution
of the self-simulating process of magnetic field penetration into
a superconductor is shown schematically in figure 6a and 6b.

\begin{center}
\includegraphics[width=2.5583in]{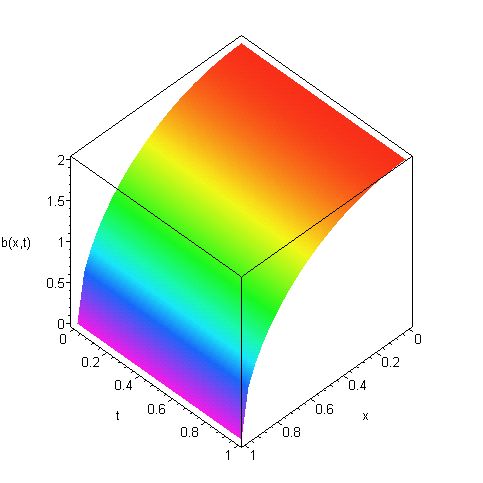}
\includegraphics[width=2.5583in]{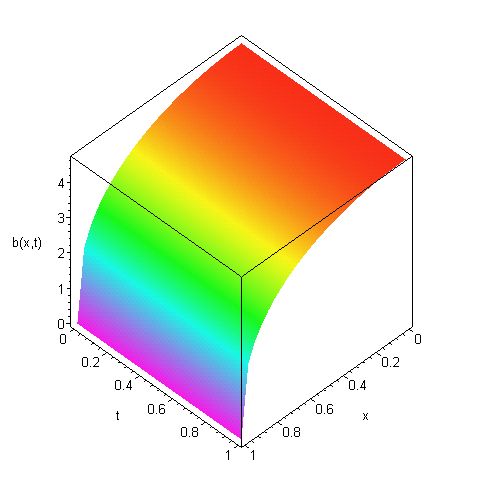}
\end{center}
\begin{center}
Fig.6a and 6b. The distributions of the normalized flux density
$b(x, t)$
at time t=1, for $\gamma$=1, 2.\\
\end{center}

 \vskip 0.5cm
\begin{center}
{\bf\S 7.2. Exponential model n=1}
\end{center}

Let us consider a solution to the nonlinear diffusion equation for
the exponential model, assuming that n=1. The nonlinear
differential equation (47), describing evolution of the magnetic
flux penetrated into the superconductor sample for the exponential
model can be transformed to the following form

\begin{equation}
\dsf{db}{dt}=\dsf{d}{dx}\left[e^b\dsf{db}{dx}\right].
\end{equation}
It is remarkable that one of explicit solution of the nonlinear
diffusion equation (60) can be obtained by using the method of
separation of variables. According to the method of separation of
variables, we look for the solution of (60) in the form

\begin{equation}
b(x,t)=a(x)g(t).
\end{equation}
Substituting this expression for b(x, t) into the partial
differential equation (60) and separating variables we obtain the
following equations for the distributions of the functions g(t)
and a(x)

\begin{equation}
e^{-\gamma g}\dsf{dg}{dt}=\lambda,
\end{equation}
\begin{equation}
\dsf{d}{dx}\left(e^{\gamma a}\dsf{da}{dx}\right)=\lambda.
\end{equation}
Integrating the differential equations (62) and (63) with respect
to t and x, respectively and using the boundary conditions we
obtain

\begin{equation}
e^{-\gamma b}=\lambda (t-t_p),
\end{equation}
\begin{equation}
e^{\gamma a}=\dsf{\lambda}{2}[(x^2-x_{p}^{2})].
\end{equation}
where $t_p$ is a constant, the parameter $x_p$ is defined by using
the integral relation (9). Combining the equations (64) and (65)
according to relation (61) we get
\begin{equation}
\phi=\dsf{1}{\gamma}\ln\left[\dsf{(x^2-x_{p}^{2})}{2(t_p-t)}\right].
\end{equation}

\vskip 0.5cm
\begin{center}
{\bf\S 8.  Magnetization}
\end{center}

Let us consider the magnetization profile of the sample which is
given by the following relation

$$
-\mu_0M(t)=b_0-\dsf{1}{d}\int_{0}^{d}b(x)dx.
$$
Substituting here the scaling law, Eq. (15) one obtains
$$
-\mu_0M(t)=b_0\left[1-\dsf{x_p(t)}{d}\psi_n\right],
$$
where d is the distance of the flux front penetrating into the
sample; $\psi_n$ is a numerical factor which is determined by the
average value of the scaling function f(z) in the region
$0<x_p<d$, so
$$
\psi_n=\dsf{1}{z_0}\int_{0}^{z_0}f(z)dz.
$$
Expressing the time dependence explicitly, the magnetization can
be written as
$$
-\mu_0M(t)=b_0\left[1-\psi_n\left(\dsf{t}{t^*}\right)^{\beta}\right],
$$
for $0<t<t^*$, where $t^*=(d/z_0)^{\beta}$.  Differentiating the
last equation and using an explicit solution for $x_p$(t) for the
case $\gamma=-1/n$ one obtains the expression to the magnetic
relaxation rate
$$
\dsf{dM}{d\ln t}\approx
\dsf{1}{n+1}\left(\dsf{t}{t^*}\right)^{1/(n+1)}.
$$
As the parameter n increases the profile of the relaxation rate is
seen to become more linear. At low temperature limit when $n\gg 1$
the relaxation rate varies linearly with time. The time dependence
of the magnetic relaxation rate is shown in figure 7.

\vskip 1cm
\begin{center}
\includegraphics[width=2.5583in]{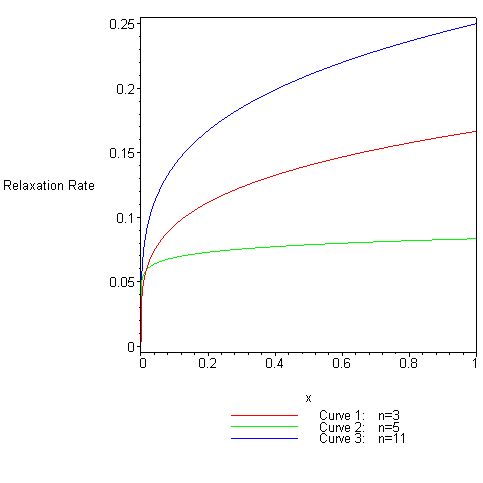}
\end{center}
\vskip 0.5cm
\begin{center}
Fig.7. The time dependence
of the magnetic relaxation rate for n=3, 5, 11.\\
\end{center}

It has been shown by Vinokur [11] that at low temperatures the
relationship between the time and relaxation rate is a linear over
a very long time period in the early stage of the flux
penetration. A such perfect linear relationship characterizes that
the system behaves a self-organized criticality.

\vskip 0.5cm
\begin{center}
{\bf Conclusion}
\end{center}

 Thus, the problem of magnetic flux penetration into the half-space
superconductor sample is studied in the flux flow regime in
parallel geometry assuming that an external magnetic field
increasing with time in accordance with the power law (7).
Assuming that the flux flow resistivity as a power-law function of
the magnetic field induction we found an exact analytical solution
for the nonlinear local magnetic flux diffusion equation. The
obtained solution describes space-time distribution of the
magnetic induction in the sample. It was shown that the magnetic
flux density profiles at different times follow the different
scaling law. Similar scaling profiles can be found for the
electric field and current density profiles, too. It was also
shown that the spatial and temporal profiles of magnetic flux
penetration in the sample depends on the set of three independent
parameters, n, q and $\alpha$. For a given parameter set n, q and
$\alpha$ the form of the scaling function b(x, t) was obtained by
solving the nonlinear diffusion equation for the magnetic field
induction analytically by a self-similar technique. We analyzed
the effect of different values of exponents on the propagation of
the flux front in the sample. Varying the parameters of the
equation, we have observed a various shapes of the magnetic flux
front. Finally, we have discussed the flux creep problem, briefly.

\vskip 0.5cm
\begin{center}
{\bf   Acknowledgements}
\end{center}

This study was supported by the NATO Reintegration Fellowship
Grant and Volkswagen Foundation Grant. We are thankful for helpful
discussions with Professor A. S. Rakhmatov. Part of the simulation
work herein was carried on in the Condensed Matter Physics at the
Abdus Salam International Centre for Theoretical Physics.

\vskip 0.5cm

\begin{center}
{\bf  References}
\end{center}

\begin{enumerate}
\item  L. D. Landau and E. M. Lifshitz, Fluid Mechanics, Pergamon,
Oxford, 1987.

\item A. A. Samarskii, V. A. Galaktionov, S. P. Kurdjumov, and A.
S. Stepanenko, Peaking Regimes for Quasilinear Parabolic
Equations, Nauka, Moskow, 1987.

\item   D. G. Aranson, J.L. Vazquez, Phys. Rev. Lett., 72, 1994.

\item   F. Bass, B. Ya. Shapiro, I. Shapiro, M. Shvartser, Physica
C 297, 269, 1998.

\item I. B. Krasnyuk, Technical Physics, 52, 2007.

\item   V. Meerovich, M. Sinder and V. Sokolovsky, Supercond. Sci.
Technol., 9, 1996.

\item   V. V. Bryksin, S. N. Dorogovstev, Physica C 215, 1993.

\item   J. Gilchrist, C. J. van der Beek, Physica C 231, 1994.

\item   J. Gilchrist, Physica C 291, 1997.

\item   D. V. Shantsev, Y. M. Galperin, and T. H. Johansen
arXiv:cond-mat/0108049 v1, 2001.

\item   V. V. Vinokur, M. V. Feigelman, and V. B. Geshkenbein,
Phys. Rev. Lett., 67, 915, 1991.

\item   Z. Koziol and E. P. Chatel, IEEE Trans. Magn., 30, 1169,
1994.

\item   W. Wang and J. Dong, Phys. Rev. B 49, 698, 1994.

\item M. Holiastou, M. Pissas, D. Niarchos, P. Haibach, U. Frey
and H. Adrian, Supercond. Sci. Technol., 11, 1998.

\item   C. P. Bean, Phys. Rev. Lett. 8, 250, 1962; Rev. Mod.
Phys., 36, 31, 1964.

\item  A. M. Campbell and J. E. Evetts, Critical Currents in
Superconductors (Taylor and Francis, London, 1972),  Moscow, 1975.

\item   P. W. Anderson , Y.B. Kim  Rev. Mod. Phys., 36. 1964.

\item   P. W. Anderson,  Phys. Rev. Lett.,  309, 317, 1962.

\item   E. Zeldov, N. M. Amer, G. Koren, A. Gupta, R. J. Gambino,
and M. W. McElfresh, Phys. Rev. Lett., 62, 3093, 1989.

\item   P. H. Kes, J. Aarts, J. van der Berg, C.J. van der Beek,
and J.A. Mydosh, Supercond. Sci. Technol., 1, 242, 1989.

\item   F. Irie and K. Yamafuji, J. Phys. Soc. Jpn., 23, 255,
1967.

\item W. A. Fietz, M. R. Beasley, J. Silcox and W. W. Webb, Phys.
Rev., 136 A335, 1964.

\end{enumerate}
}\end{multicols}

\end{document}